\documentclass[graybox]{svmult}
\usepackage{mathptmx}       
\usepackage{helvet}         
\usepackage{courier}        
\usepackage{type1cm}        
\usepackage{makeidx}         
\usepackage{graphicx}        
\usepackage{multicol}        
\usepackage[bottom]{footmisc}

\makeindex             

\begin{document}

\title*{Highly Charged Ions in Rare Earth Permanent Magnet Penning Traps}
\author{Nicholas D. Guise, Samuel M. Brewer, and Joseph N. Tan}
\institute{Nicholas D. Guise \at National Institute of Standards and Technology, Gaithersburg, MD 20899, USA\newline
\email{nicholas.guise@nist.gov}
\and Samuel M. Brewer \at University of Maryland, College Park, MD, 20742, USA
\and Joseph N. Tan \at National Institute of Standards and Technology, Gaithersburg, MD 20899, USA}
%
%
\maketitle

\abstract{A newly constructed apparatus at the United States National Institute of Standards and Technology (NIST) is designed for the isolation, manipulation, and study of highly charged ions.  Highly charged ions are produced in the NIST electron-beam ion trap (EBIT), extracted through a beamline that selects a single mass/charge species, then captured in a compact Penning trap.  The magnetic field of the trap is generated by cylindrical NdFeB permanent magnets integrated into its electrodes.  In a room-temperature prototype trap with a single NdFeB magnet, species including Ne$^{10+}$ and N$^{7+}$ were confined with storage times of order 1 second, showing the potential of this setup for manipulation and spectroscopy of highly charged ions in a controlled environment.  Ion capture has since been demonstrated with similar storage times in a more-elaborate Penning trap that integrates two coaxial NdFeB magnets for improved $B$-field homogeneity.  Ongoing experiments utilize a second-generation apparatus that incorporates this two-magnet Penning trap along with a fast time-of-flight MCP detector capable of resolving the charge-state evolution of trapped ions.  Holes in the two-magnet Penning trap ring electrode allow for optical and atomic beam access.  Possible applications include spectroscopic studies of one-electron ions in Rydberg states, as well as highly charged ions of interest in atomic physics, metrology, astrophysics, and plasma diagnostics.}

\section{BACKGROUND}
\label{sec:background}

A Penning trap combines a quadrupole electric potential with a strong magnetic field in order to trap a charged particle in three dimensions.  The magnetic field provides radial confinement while the electric field provides confinement along the axis of the magnetic field.  Penning traps see wide application in precision physics experiments ranging from measurements of fundamental constants \cite{Hanneke2008} to quantum information studies \cite{Biercuk2009}.

The use of Penning traps to recapture ions from accelerator or electron-beam ion trap (EBIT) sources has drawn longstanding interest at various facilities, as a means of extending the reach of spectroscopic studies with exotic ions/atoms and highly charged ions (HCI).  Examples include antimatter studies \cite{Andresen2011, Gabrielse2008, Gabrielse1986}, mass spectrometry and nuclear research at accelerator facilities \cite{Kluge2003}, early experiments recapturing HCI from an EBIT source \cite{Schneider1994}, and ongoing efforts in HCI laser spectroscopy \cite{Andjelkovic2010} and mass measurement \cite{Simon2011, Repp2012}.  The typical high-precision Penning trap utilized in these experiments requires a superconducting solenoid to generate its axial magnetic field of several Tesla.  The associated cryostat makes a high-field Penning trap a large and expensive apparatus, similar in scale to an EBIT itself (see Sec. \ref{sec:EBIT}).  A lower-field Penning trap, configured in similar geometry but using a room-temperature solenoid, may function as a pre-cooling stage; for example, evaporative cooling of Ar$^{16+}$ ions was recently demonstrated in a 1.1 Tesla trap within the SMILETRAP II mass spectrometer \cite{Hobein2011}.

An alternative Penning trap design relies upon rare-earth compounds such as Nd$_2$Fe$_{14}$B (NdFeB) to generate the trapping magnetic field.  While the magnetic field generated by rare earth magnets is typically less than 1 Tesla in the trapping region, the rare earth magnets are relatively inexpensive and allow increased flexibility in trap design.  In earlier work, a compact Penning trap with its magnetic field generated by radially-oriented NdFeB wedges (assembled in two rings) was used to confine light ions loaded within a 77 K apparatus \cite{Gomer1995}; another trap that combined axially oriented magnets with epoxy-bonded arrays of radially-magnetized segments was used to store molecular anions at room temperature \cite{Suess2002}.

To facilitate operation at the high voltages necessary for catching highly charged ions from the NIST EBIT, we have developed two new compact Penning traps utilizing only cylindrical, axially-oriented NdFeB magnets in robust unitary architectures, with the trap electrodes and magnetic elements integrated as a unit.  These traps are designed to provide easy access for laser and atomic beam manipulation of trapped ions, and to satisfy space constraints at the NIST EBIT facility.  Here we report initial studies with highly charged ions produced in the EBIT, including bare nuclei, which are extracted and captured in these extremely compact Penning traps.

The primary goal of our experimental effort is to enable precise spectroscopic studies of one-electron (H-like) ions in circular states, synthesized from stored bare nuclei.  Theoretical studies at NIST \cite{Jentschura2008, Jentschura2010} have shown that high angular momentum states of one-electron ions are better understood than low-lying states; in some cases, the energy level prediction for high-L states attains accuracy comparable to the precision of optical frequency combs.  This offers a new avenue for the determination of fundamental constants, such as the Rydberg constant, if precise optical measurements can be realized to test theory.  Such tests could potentially be useful in making improved measurements of the constants of nature that form the basis of modern metrology \cite{Mohr2008a, Mohr2008b}---efforts that are more compelling in the wake of the large discrepancy in proton radius determinations that resulted from recent muonic hydrogen Lamb-shift measurements \cite{Pohl2010, Nez2011}.

The ion-trapping techniques under development may also prove useful for other studies of highly charged ions.  One example application (see Sec. \ref{sec:Summary}) is measuring the lifetime of certain metastable states, where most previous studies have been performed inside an EBIT environment \cite{Lapierre2006} or in Kingdon-type ion traps \cite{Church1999}.

A brief outline of this paper follows:
\begin{itemize}
\item{Section \ref{sec:expapp} reviews Penning trap basics, provides details of the compact NdFeB trap designs, and describes HCI extraction from the NIST EBIT.}
\item{Section \ref{sec:expresults} presents initial results from HCI capture and storage.}
\item{Section \ref{sec:Summary} discusses potential applications and future directions for this work.}
\end{itemize}

\section{Experimental Apparatus}
\label{sec:expapp}

\subsection{Compact Penning Traps with Unitary Architecture}
\label{subsec:compactpenning}

\begin{figure}[tp]
\includegraphics[width=\textwidth]{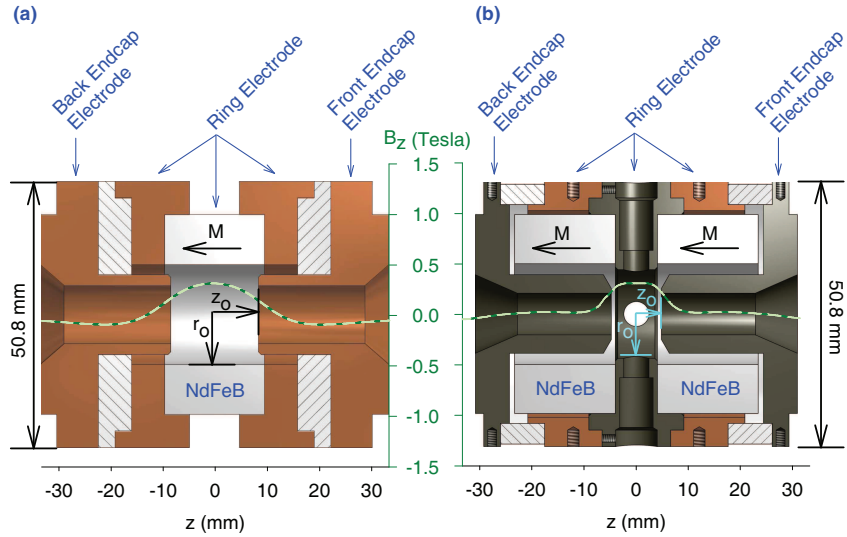}
\caption{Cross-sectional views of (a) one-magnet; and (b) two-magnet compact NdFeB Penning traps.  The cylindrical symmetry axis is horizontal.  Vector \textbf{M} indicates the direction of remnant magnetization for each NdFeB magnet.  The on-axis magnetic field profile for each trap is plotted as an overlay (green dashed line).  Detailed field profiles are provided in a separate publication \cite{Tan2012}.}
\label{fig:TrapSchematics}       
\end{figure}

A basic cylindrical Penning trap (Fig.  \ref{fig:TrapSchematics}) consists of a ring electrode between two endcap electrodes.  A near-quadrupole potential is created by applying voltage difference $\Delta V$ between the ring and the endcaps, forming a potential well along the $z$-axis but a potential hill in the radial dimension $r$.  A magnetic field $\mathbf{B}$ applied along the $z$-axis provides radial confinement, completing the trap.  In the canonical treatment of a single ion in a Penning trap \cite{Brown1986}, the trapped ion motion is described by a superposition of three oscillations: (1) a simple harmonic axial motion, at the ``axial frequency'' $2 \pi \omega_z$ ; (2) a high-frequency cyclotron orbit around the axis of the magnetic field, at the trap-modified ``cyclotron frequency'' $2 \pi \omega'_c$ ; and (3) a slower radial motion due to $\mathbf{E} \times \mathbf{B}$ drift, at the ``magnetron frequency'' $2 \pi \omega_m$.  These frequencies typically observe the hierarchy $\omega'_c >> \omega_z >> \omega_m $.

The axial frequency is defined by
\begin{equation}
\label{eq:axialfreq}
\omega_z^2 = \lambda \frac{q \Delta V}{m d^2} ,
\end{equation}
where $q$ is the ion charge, $m$ is the ion mass, and parameters $d$ and $\lambda$ depend on trap geometry: $d$ is an effective trap dimension given by
\begin{equation}
\label{eq:trapd}
d^2 = \frac{1}{2} \left( z_0^2+r_0^2/2 \right) ,
\end{equation}
while $\lambda$ describes the degree to which the electrodes produce a quadrupole potential; $\lambda \approx 1$ for near-hyperbolic surfaces and small-amplitude oscillations.  The distances $z_0$ and $r_0$ are shown in Fig.  \ref{fig:TrapSchematics}.

The free-space cyclotron frequency in a magnetic field of magnitude $B$ is given by
\begin{equation}
\label{eq:freespacecyc}
\omega_c = \frac{qB}{m} .
\end{equation}
The presence of the radial electrostatic potential in a Penning trap shifts this cyclotron frequency to its trap-modified value,
\begin{equation}
\label{eq:trapcyc}
\omega'_c = \omega_c-\omega_m ,
\end{equation}
where $\omega_m$ is the frequency of the magnetron motion,
\begin{equation}
\label{eq:magfreq}
\omega_m = \frac{1}{2} \left( \omega_c- \sqrt{\omega_c^2-2 \omega_z^2} \right).
\end{equation}
To maintain localized orbits, the magnetron frequency must be real, hence
\begin{equation}
\label{eq:stabilitycrit}
\omega_c^2-2 \omega_z^2 > 0
\end{equation}
is a confinement criterion for Penning traps.

We present two designs for compact Penning traps utilizing NdFeB permanent magnets in a unitary architecture.  Table \ref{tab:trapfreqs} provides frequencies of motion for various HCI species of interest, in the one-magnet (Sec.  \ref{sec:OneMagTrap}) and two-magnet (Sec.  \ref{sec:TwoMagTrap}) trap designs.  Values of parameter $\lambda$ (Eqn.  \ref{eq:axialfreq}) are determined by numerical simulation.  Frequencies are calculated using Eqns.  \ref{eq:axialfreq}-\ref{eq:magfreq} for the simple case of a single ion oscillating near trap center.  For ion clouds of significantly larger number or physical size, these frequencies are expected to shift due to trap anharmonicities and magnetic field gradients---precise trap frequencies in this regime may be obtained through numerical simulation.

\subsubsection{One-Magnet Penning Trap}
\label{sec:OneMagTrap}
Our simplest trap design (Fig.  \ref{fig:TrapSchematics}a) uses a single NdFeB magnet as the ring electrode in an uncompensated open-endcap Penning trap.  The NdFeB magnet is sandwiched between oxygen-free high-conductivity (OFHC) copper rings through which we apply the ring bias voltage.  The trap endcap electrodes are machined from OFHC copper.  MACOR ceramic glass spacers provide electrical isolation between neighboring electrodes.  Characteristic trap dimensions for the one magnet trap are $r_o=$ 9.53 mm and $z_o =$ 8.39 mm.  The value of parameter $\lambda$ (Eqn.  \ref{eq:axialfreq}) is 0.814.  Typical oscillation frequencies for representative ion species are provided in Table \ref{tab:trapfreqs}.

\subsubsection{Two-Magnet Penning Trap}
\label{sec:TwoMagTrap}
A design with two NdFeB magnets (Fig.  \ref{fig:TrapSchematics}b) increases the complexity of assembly but yields several advantages over the one-magnet design.  Positioning the magnets behind electrical iron endcaps serves to minimize the effect of physical and magnetic imperfections, and the gap between the magnets now permits holes in the $z = 0$ plane, for laser or atomic beam access into the trapping region.  Our two-magnet trap design also has improved $B$-field homogeneity in the range $z = 2$ mm to $z = -2$ mm, compared to the one-magnet design \cite{Tan2012}.

Characteristic trap dimensions for the two-magnet trap are $r_0 =$ 8.50 mm and $z_0 =$ 4.74 mm.  The two NdFeB magnets dovetail onto an electrical iron yoke and two OFHC copper rings, together comprising the ring electrode.  The iron yoke has four evenly spaced radial holes, one of which contains an aspheric lens for light collection.  The endcap electrodes are machined from electrical iron.  MACOR spacers are again used to provide electrical isolation between neighboring electrodes.  The value of parameter $\lambda$ (Eqn.  \ref{eq:axialfreq}) is 0.854.  Typical oscillation frequencies for representative ion species are provided in Table \ref{tab:trapfreqs}.
%
\begin{table}
\caption{Approximate motional frequencies $2 \pi \omega_z$, $2 \pi \omega'_c$, and $2 \pi \omega_m$ for various highly charged ions confined in the compact Penning trap designs of Fig.  \ref{fig:TrapSchematics}.  The applied trapping well is $\Delta V=$ 10 V, and the magnetic field has $B_z=$0.3 T.  Calculated values assume a single trapped ion in the limit of small-amplitude oscillations.}
\label{tab:trapfreqs}       
\begin{tabular}{p{1.5cm}p{1.5cm}p{1.5cm}p{1.5cm}p{0.25cm}p{1.5cm}p{1.5cm}p{1.5cm}}
\hline\noalign{\smallskip}
&\multicolumn{3}{c}{one-magnet trap, $\lambda=0.814$}&&\multicolumn{3}{c}{two-magnet trap, $\lambda=0.854$}\\
\hline\noalign{\smallskip}
Ion&Axial\hspace{5 mm}Freq.  (kHz)&Cyclotron Freq.  (kHz)&Magnetron Freq.  (kHz)&&Axial\hspace{5 mm}Freq.  (kHz)&Cyclotron Freq.  (kHz)&Magnetron Freq.  (kHz)\\
\noalign{\smallskip}\svhline\noalign{\smallskip}
$^{20}$Ne$^{10+}$&415&2266&38.0&&597&2224&80.1 \\ \hline\noalign{\smallskip}
$^{20}$Ne$^{9+}$&393&2036&38.0&&566&1994&80.5 \\ \hline\noalign{\smallskip}
$^{20}$Ne$^{8+}$&371&1805&38.1&&534&1763&80.9 \\ \hline\noalign{\smallskip}
$^{40}$Ar$^{16+}$&371&1806&38.1&&534&1764&80.9 \\ \hline\noalign{\smallskip}
$^{40}$Ar$^{15+}$&359&1691&38.2&&517&1648&81.2 \\ \hline\noalign{\smallskip}
$^{40}$Ar$^{14+}$&347&1576&38.2&&500&1533&81.5 \\ \hline\noalign{\smallskip}
$^{40}$Ar$^{13+}$&334&1460&38.3&&481&1417&81.8 \\ \hline
\noalign{\smallskip}\hline\noalign{\smallskip}
\end{tabular}
\end{table}
\subsection{Production and Extraction of Highly Charged Ions}
\label{sec:EBIT}
The electron-beam ion trap (EBIT) is a laboratory device for producing highly charged ions of various species.  Invented in the late 1980s \cite{Levine1988}, EBITs now appear in multiple facilities worldwide.  The NIST EBIT \cite{Gillaspy1997}, in operation since 1993, incorporates an electron gun capable of generating a beam of current larger than 100 mA within an ultrahigh vacuum enclosure.  The nearly mono-energetic electron beam is accelerated through high voltage, reaching kinetic energies up to 30 keV as it travels through the central axis of a stack of cylindrical drift tubes.  Superconducting Helmholtz coils generate a strong axial magnetic field ($\approx 3$ Tesla) which compresses the radial extent of the electron beam, producing an extremely high current density.  Highly charged ions are produced through repeated electron-impact ionization of an injected gas.  These positively-charged ions are trapped in a configuration that resembles an open-endcap Penning trap, with the cylindrical drift tubes biased to form an electrostatic trapping potential along the axis, but with the highly compressed electron beam also providing radial confinement.
\begin{figure}[hpb]
\includegraphics[width=\textwidth]{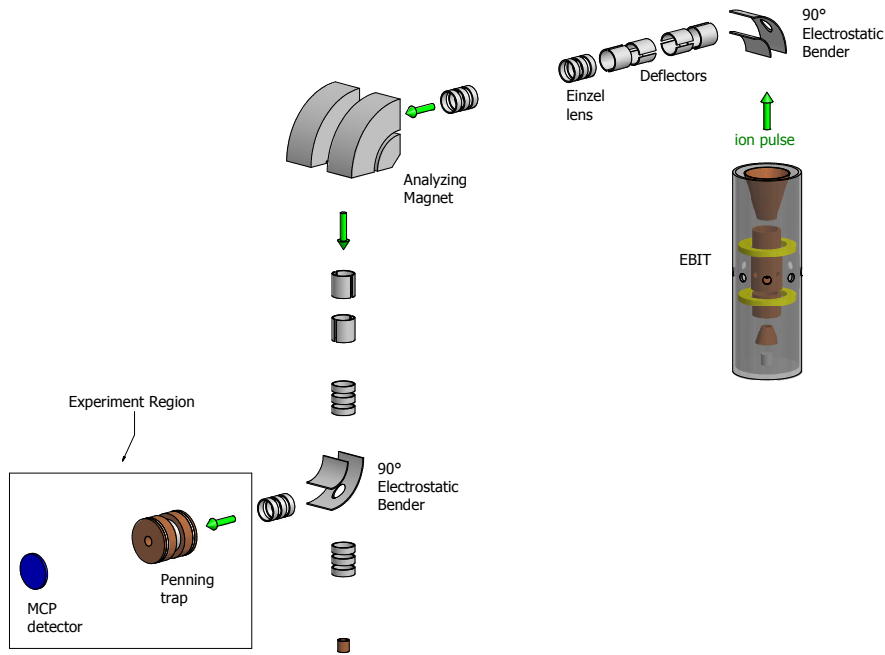}
\caption{Cartoon of extracted ion path from EBIT to the compact Penning trap.  Details of the experiment region
are shown in Fig.  \ref{fig:ExptBeamline}.  Details of EBIT beamline elements are as previously published \cite{Ratliff1998}.}
\label{fig:EBITSchematic}       
\end{figure}

The EBIT produces a mixture of several charge states, determined by various tuning parameters and by the available ionizing energy of the electron beam.  This HCI mixture is extracted by quickly ramping the EBIT middle drift tube up in voltage.  As the pulse of ions accelerates out of the drift tube region, each ion acquires substantial kinetic energy, roughly equal to the ion charge times the effective EBIT operating voltage.  Typical energies for the experiments described below are between 2 keV to 4 keV per unit charge.  In the following sections we describe how this ion pulse is controlled and recaptured into a Penning trap made extremely compact with a unitary architecture.

\subsubsection{EBIT Extraction Beamline} %
\begin{figure}[tp]
\includegraphics[width=\textwidth]{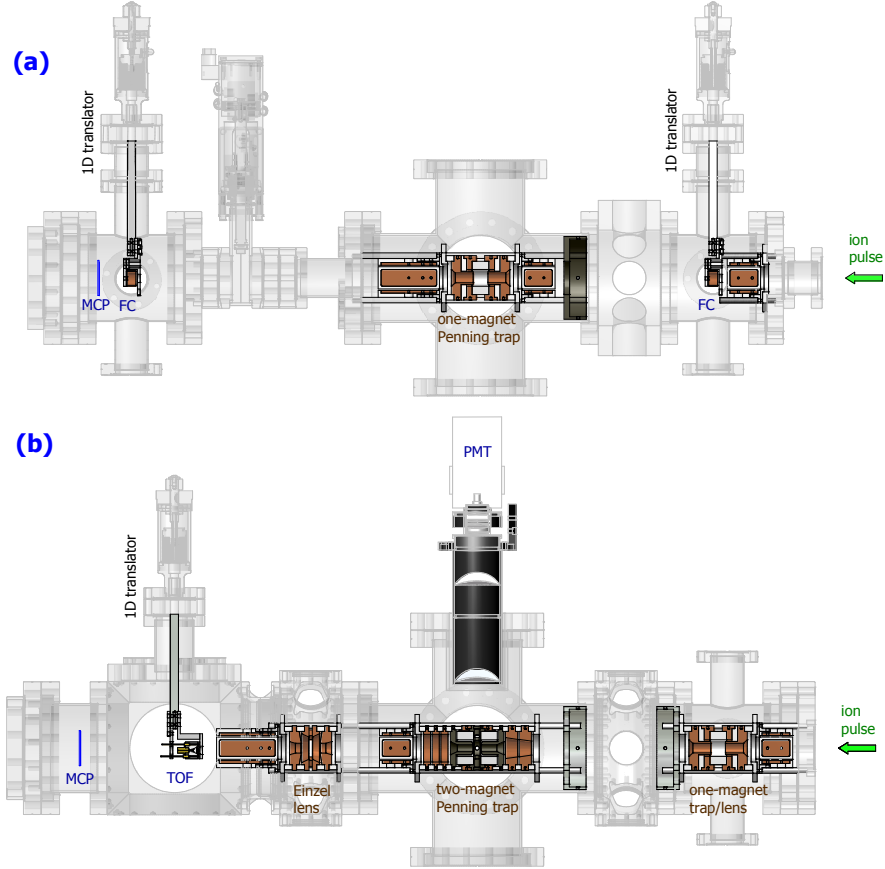}
\caption{Detailed view of the experiment region for two different Penning trap configurations: (a) Beamline incorporating the one-magnet Penning trap (Fig. \ref{fig:TrapSchematics}a) and three sets of steering plates; detection elements include retractable Faraday cups (FC) and a position-sensitive microchannel plate (MCP).  (b) Beamline incorporating the two-magnet Penning trap (Fig. \ref{fig:TrapSchematics}b), with additional focusing elements, a retractable time-of-flight MCP detector (TOF), and visual access to a photomultiplier tube (PMT).}
\label{fig:ExptBeamline}       
\end{figure}

Figures \ref{fig:EBITSchematic} and \ref{fig:ExptBeamline} provide a schematic overview of the experiment.  The approximate path length from the EBIT to the Penning trap capture/detection region is 7.2 meters.  Specifics of the NIST EBIT extraction beamline are provided elsewhere \cite{Ratliff1998}.

The various tuning and focusing elements in the extraction beamline nominally maintain ion kinetic energy, and all but the analyzing magnet act equally on ions of different mass/charge.  The analyzing magnet allows selection of a particular ion charge state.  At the proper analyzing magnetic field value for a given mass/charge ratio, the ion is deflected by the necessary amount to continue through the beamline; it then passes through the Penning trap and strikes a Faraday cup or microchannel plate (MCP) for detection.

\begin{figure}[tp]
\includegraphics[width=\textwidth]{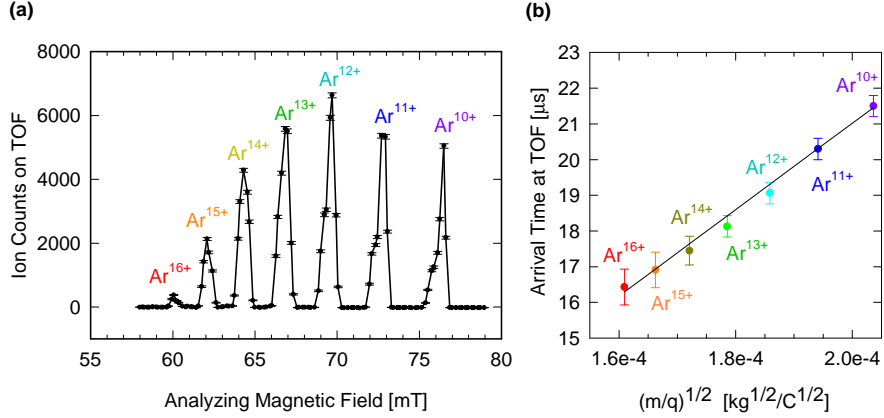}
\caption{(a) Typical ``mass scan'' of argon ions extracted from the EBIT with $V_0 \approx 2.85$ kV (see Eqn.  \ref{eq:timeofflight}).  The number of ions is counted by measuring the signal on a time-of-flight MCP detector as the analyzing magnetic field is varied to select out different mass/charge states.  (b) Arrival times of various argon ions at the TOF detector, relative to the EBIT dump trigger.  Error bars represent 1$\sigma$ uncertainty.}
\label{fig:MassScan}       
\end{figure}

Figure \ref{fig:MassScan} shows the result of a "mass scan" performed by sweeping the analyzing magnetic field and tracking the detected ion signal on the time-of-flight MCP.  Charge states are identified from the spacing between peaks and the known ion kinetic energy.  MCP signals may then be adjusted for the charge state of each ion to extract the actual number of each charge state striking the detector (Fig.  \ref{fig:MassScan}a).  The assignment of charge states is confirmed by plotting the arrival time of each ion at the TOF detector, measured as a delay from the EBIT dump trigger (Fig.  \ref{fig:MassScan}b).  This arrival time, dominated by the transit time through the EBIT extraction beamline, shows the expected linear dependence on the square root of ion mass/charge ratio, for lossless motion in conservative electric potentials.  In particular, the time $t$ required for an ion to traverse the extraction beamline of length $L$ is given by
\begin{equation}
\label{eq:timeofflight}
t=\int_0^L \! \frac{dx}{\mathrm{v}(x)}=\sqrt{\frac{m}{q}}\int_0^L \! \frac{dx}{\sqrt{2(V_0-V(x))}} \, ,
\end{equation}
where the $x$-coordinate is directed along the beamline, $\mathrm{v}(x)$ is the ion velocity, $V(x)$ is the electric potential at position $x$ (except near deflectors or focusing elements, $V(x)\approx0$), and $V_0$ is the effective operating voltage of the EBIT, i.e.  the initial kinetic energy of an extracted ion is $qV_0$.

\begin{figure}[tp]
\includegraphics[width=\textwidth]{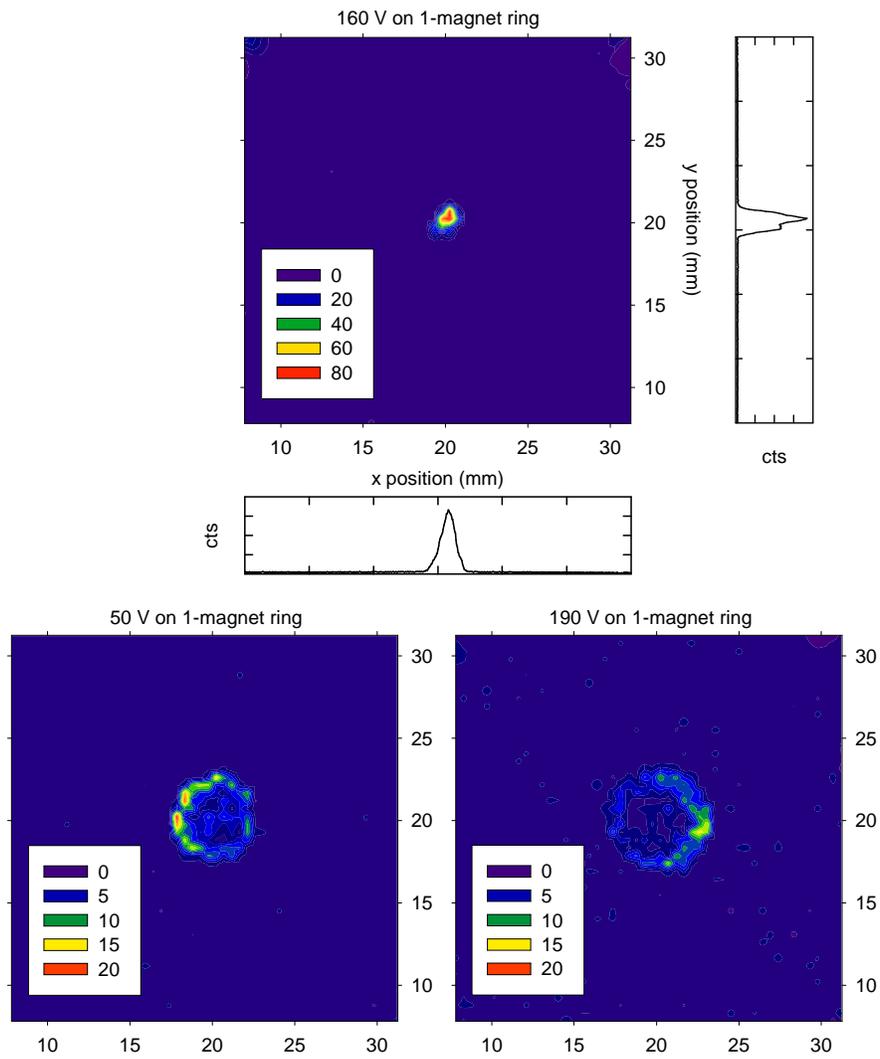}
\caption{Beam spots observed on the XY position-sensitive MCP detector.  Ne$^{10+}$ ions passing through the one-magnet experiment beamline (Fig.  \ref{fig:ExptBeamline}a) are focused by applying different tuning potentials to the one-magnet Penning trap, which functions as an Einzel lens.  Colors indicate number of ions.}
\label{fig:BeamSpots}       
\end{figure}

After setting the analyzing magnetic field to isolate the charge state of interest, we can use steering and focusing elements to optimize the beam image on the position-sensitive MCP detector.  Figure \ref{fig:BeamSpots} shows representative beam spot images taken while operating the one-magnet Penning trap as an Einzel lens for focusing---this is the primary function of the one-magnet trap in our two-magnet experiment configuration (Fig.  \ref{fig:ExptBeamline}b).  The ion beam is centered by applying correction voltages to three sets of orthogonal steering plates, apertured through the trap electrodes, and focused to a spot as small as $\approx 3$ mm diameter on the detector.
\subsubsection{Capture and Detection of Extracted Ions}
To capture ions extracted from the EBIT, the Penning trap electrodes are biased in a step configuration, with the front endcap "low," the back endcap "high," and the ring electrode at an intermediate voltage.  Each electrode then receives an additional constant high voltage bias, roughly equal to the effective EBIT operating voltage, in order to match the kinetic energy of incoming ions.   As the ions approach the Penning trap, they climb the potential hill and slow down; once inside the Penning trap region, the ions are then captured by rapidly increasing the front endcap voltage to match the back endcap voltage, closing the trap and creating a local potential well centered on the ring electrode.  Once captured, ions may be held for several seconds with the trap voltages in this storage configuration, then dumped to a detector by lowering the back endcap voltage.  High voltage pulsers, capable of rise times $<$100 ns, are used to accomplish the fast switching between voltage settings for ion capture, storage, and dumping.
\begin{figure}[tp]
\includegraphics[width=\textwidth]{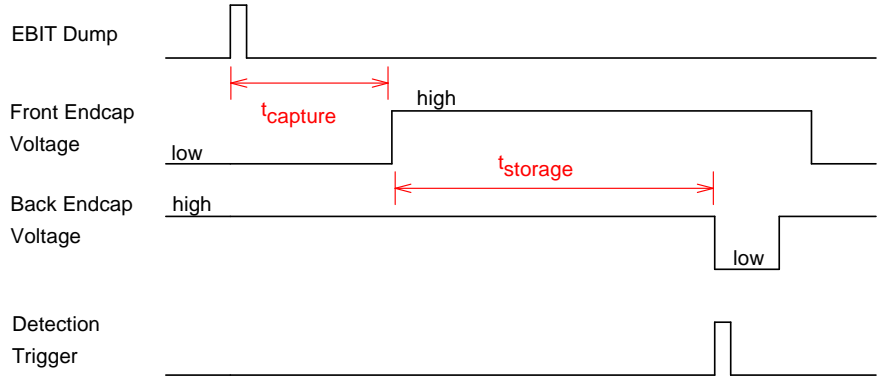}
\caption{Simplified timing diagram for ion capture, storage, and detection.}
\label{fig:Timing}       
\end{figure}

A simplified timing diagram for the experiment is shown in Fig.  \ref{fig:Timing}.  Ions are extracted from the EBIT by pulsing the middle drift tube up in voltage.  After waiting some amount of time (t$_\mathrm{capture}$) for the extracted ions to transit from the EBIT, the front endcap then pulses up in voltage to capture ions just after they arrive in the Penning trap region.  After an additional storage time (t$_\mathrm{storage}$), the Penning trap back endcap pulses down in voltage to dump the ions towards the MCP detector.  A trigger pulse simultaneous with the back endcap pulser allows synchronization of the detection electronics.

\begin{figure}[tp]
\includegraphics[width=\textwidth]{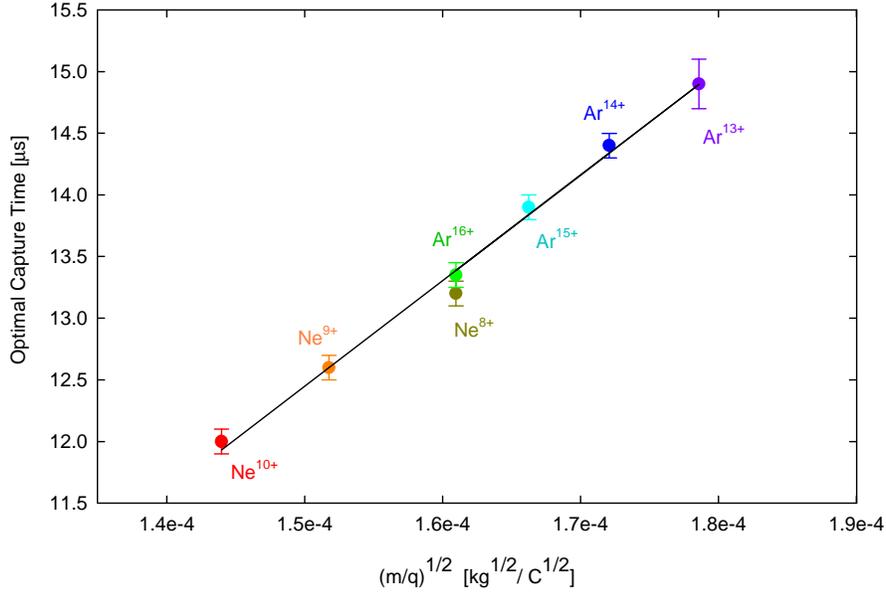}
\caption{Optimal capture time (t$_{\mathrm{capture}}$ in Fig.  \ref{fig:Timing}) for various ion species in the two-magnet Penning trap.  The effective EBIT extraction voltage ($V_0$ in Eqn.  \ref{eq:timeofflight}) is $\approx 4.3$ kV.  Error bars represent 1$\sigma$ uncertainty.}
\label{fig:CaptureTimes}       
\end{figure}

The Penning trap front endcap must pulse closed at precisely the right time in order to catch any ions.  If the trap closes too soon, the ions will not yet have reached the trapping region; if too late, they will have already exited, having been turned around by the high back endcap potential.   Experimentally we find that timing resolution of order 10 ns is sufficient to optimize the ``capture time'' at which the trap is closed.  To determine the best capture settings for a given ion species, we set the storage time to a short and fixed value (typically 1 ms), then sweep the capture time to optimize the number of detected ions.  Results of this optimization are shown in Fig.  \ref{fig:CaptureTimes} for various ion species captured in the two-magnet Penning trap.  As in Fig.  \ref{fig:MassScan}b, this is essentially a measure of transit time from the EBIT to the Penning trap, and we observe the expected linear dependence on the square root of ion mass/charge ratio (Eqn.  \ref{eq:timeofflight}).

\section{Experimental Results}
\label{sec:expresults}
We have demonstrated the effectiveness of compact Penning traps with unitary architecture for capture and storage of highly charged ions, using both trap designs presented in Fig.  \ref{fig:TrapSchematics}.  Initial experiments with highly charged ions extracted from the NIST EBIT utilized the one-magnet trap and the experiment beamline shown in Fig.  \ref{fig:ExptBeamline}a.  Recent experiments utilize the two-magnet trap and the more complex setup of Fig.  \ref{fig:ExptBeamline}b.  Capture and detection follows the scheme described in Fig.  \ref{fig:Timing}.  Sample results are presented below.

\subsection{One-Magnet Penning Trap with Position-Sensitive MCP Detector}
\label{sec:OneMagData}
Ion species captured in the one-magnet Penning trap include Ne$^{10+}$, Ne$^{9+}$, Ne$^{8+}$, Ar$^{16+}$, Ar$^{15+}$, Ar$^{14+}$, Ar$^{13+}$, and N$^{7+}$.  The only means of ion detection in the one-magnet experiment configuration (Fig.  \ref{fig:ExptBeamline}a) is a position-sensitive MCP detector.  This detector has an active area of diameter 40 mm, with a resistive anode encoder that correlates each ion count event with a position of impact on the MCP.   While useful for beam tuning (e.g.  Fig.  \ref{fig:BeamSpots}), this encoding has the unwanted side-effect of relatively slow response time, of order 1 $\mu$s.  To avoid dead-time errors generated by large ion count rate, we use a purposefully inefficient dump scheme.  When dumping ions from the Penning trap, the back endcap voltage is set equal to or slightly higher than the ring voltage.  Without a strong potential gradient across the trap, many ions escape to the electrode walls, but a fraction of the ions spill out towards the detector.  For trap storage times longer than 0.5 seconds, this scheme results in ion numbers low enough to be individually counted without evidence of pileup on the position-sensitive detector.
\begin{figure}[tp]
\includegraphics[width=\textwidth]{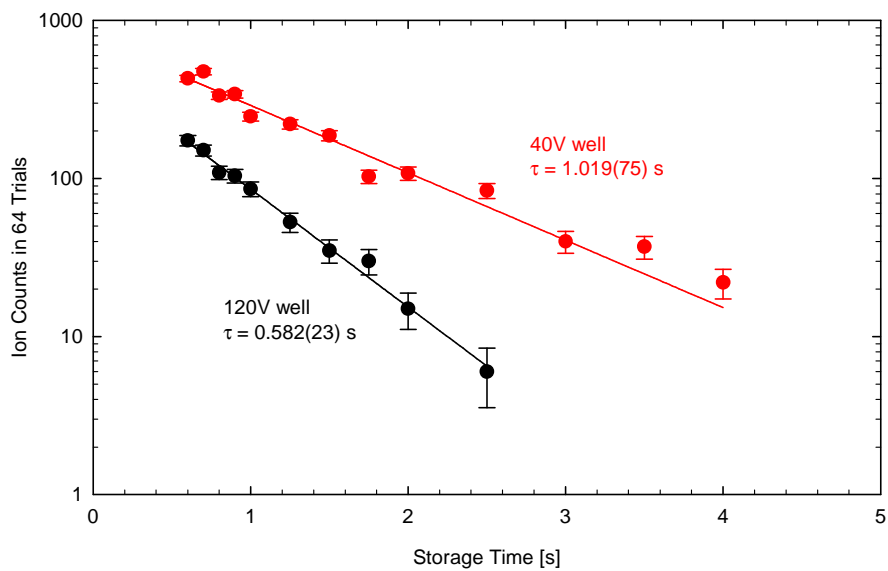}
\caption{Storage of Ne$^{10+}$ ions in the one-magnet Penning trap.  The number of ions detected on the position-sensitive MCP detector is plotted as a function of storage time, with applied potential differences of $\Delta V = 40$ V (red) and $\Delta V =$ 120 V (black) between ring and endcap electrodes.  Error bars represent 1$\sigma$ uncertainty.}
\label{fig:OneMagStorage}       
\end{figure}

Figure \ref{fig:OneMagStorage} shows the result of storage time measurements in the one-magnet Penning trap.  The ``low'' voltage applied to the back endcap electrode while dumping the trap is set nominally equal to the ring voltage, intentionally reducing detected ion counts as described above.  Storage times are measured by fitting a single exponential decay curve to the data.  Residual gas pressure in the trapping chamber was roughly $2.4 \times 10^{-7}$ Pa ($1.8 \times 10^{-9}$ Torr).

The one-magnet trap beamline (Fig.  \ref{fig:ExptBeamline}a) is intended primarily as a proof-of-concept prototype; the position-sensitive detector confirms the capture of HCI but cannot count large ion fluxes nor distinguish between different charge states.  However, the storage times observed in Figure \ref{fig:OneMagStorage}, accounting for higher background gas pressure, are comparable to those observed in the two-magnet trap for the sum of all charge states (Fig.  \ref{fig:TwoMagStorage}c, Sec.  \ref{sec:TwoMagData}).  The simplicity of the one-magnet design makes it an attractive option as a basic ion trap or focusing element; addition of a TOF detector (Sec.  \ref{sec:TwoMagData}), for example, would enable further studies of charge-exchange and trap dynamics.

\subsection{Two-Magnet Penning Trap with Time-of-Flight (TOF) MCP Detector}
\label{sec:TwoMagData}
Ion species captured in the two-magnet Penning trap include Ne$^{10+}$, Ne$^{9+}$, Ne$^{8+}$, Ar$^{16+}$, Ar$^{15+}$, Ar$^{14+}$, Ar$^{13+}$, and Kr$^{17+}$.  The two-magnet experiment beamline (Fig.  \ref{fig:ExptBeamline}b) incorporates the one-magnet trap as a focusing element and contains two additional detection options compared to the one-magnet trap beamline.  First, trapped ions may be detected optically via a photomultiplier tube (Sec.  \ref{sec:Summary}).  Second, a time-of-flight (TOF) MCP detector, mounted on a one-dimensional translator, may be inserted into the beamline in front of the position-sensitive detector.  The TOF detector has active diameter of 8 mm and response time of order 1 ns.  High voltage bias across the TOF MCP selects one of two operating modes: pulse-counting or analog amplification.  Our experiments generally utilize the analog mode, in which the TOF detector functions as a fast current amplifier with gain of $10^5$ to $10^6$.  Ion counts are extracted from the TOF signal strength by accounting for the known charge state of the incident ion.  

\begin{figure}[tp]
\includegraphics[width=\textwidth]{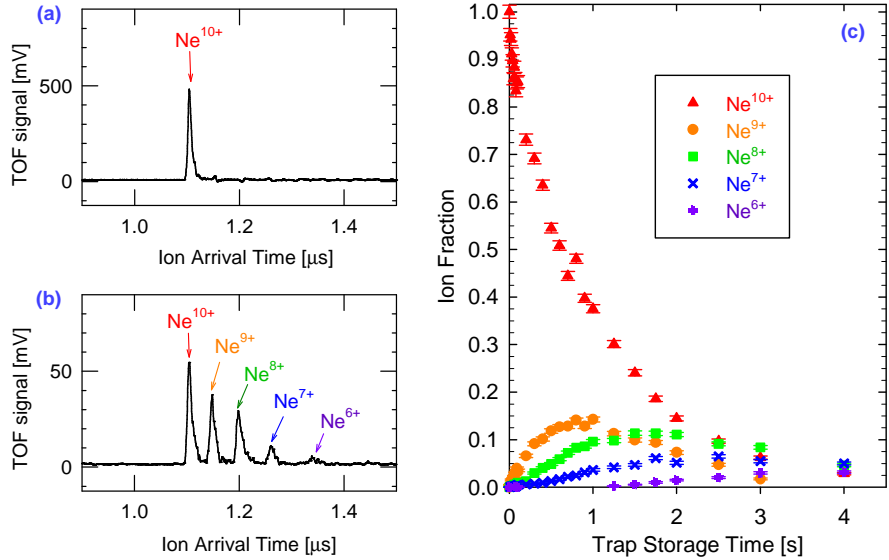}
\caption{Detection of highly charged ions in the two-magnet Penning trap, demonstrating charge-state resolution with the time-of-flight MCP detector (TOF).  Neon ions are confined in the trap for storage times ranging from 1 ms to 4 s, with applied potential difference $\Delta V =$ 10 V between the ring and endcap electrodes.  Output of the TOF detector versus arrival time relative to the Penning trap dump trigger is shown for storage times of (a) 1 ms and (b) 2 s.  The detector signal scale is magnified by 10x from (a) to (b).  The TOF signal peak for each charge state is converted to ion counts, and the evolution of charge states via charge exchange is shown in (c), normalized to the initial population of Ne$^{10+}$ ions observed at the 1 ms storage time.  Error bars represent 1$\sigma$ uncertainty.}
\label{fig:TwoMagStorage}       
\end{figure}

Figure \ref{fig:TwoMagStorage} shows ion storage measurements in the two-magnet Penning trap.  The back endcap dump voltage is set to 400 V below the ring voltage, ramping the ions out in a narrow pulse; since the TOF detector is capable of much higher ion count rates than the position-sensitive detector, pileup errors (Sec.  \ref{sec:OneMagData}) are not observed even for storage times as short as $\approx$ 1 ms.  Time resolution of the TOF detector is sufficiently fast to permit the observation of ion charge-exchange processes.  The ion species initially loaded in the two-magnet trap undergoes charge-exchange collisions with residual background gas.  The product ion species will in general still oscillate with frequencies that satisfy the Penning trap confinement criterion (Eqn.  \ref{eq:stabilitycrit}, Table \ref{tab:trapfreqs}), hence it will remain trapped and can continue to charge-exchange down to lower charge states.  Fig.  \ref{fig:TwoMagStorage}c shows the evolution of lower charge states as a function of storage time in the two-magnet trap.  The charge-exchange products are detected as separate peaks in the TOF detector signal (Fig.  \ref{fig:TwoMagStorage}a-b), delayed from the initial ion peak due to higher mass/charge.

Residual background gas pressure for these measurements is 1.7 $\times 10^{-7}$ Pa ($1.3 \times 10^{-9}$ Torr).  We observe an exponential decay in the number of ions detected on the TOF MCP detector as a function of trap storage time: the initial charge state (Ne$^{10+}$; red triangles in Fig.  \ref{fig:TwoMagStorage}c) decays with time constant 1.09(2) s, and the sum of all charge states decays with time constant 2.41(6) s.  Decay rates of both the initial charge state (dominated by charge-exchange), and of the sum of all charge states (dominated by ion cloud expansion/loss), are observed to increase linearly with background gas pressure \cite{Tan2012}.

\section{Summary and Outlook}
\label{sec:Summary}
We have developed compact Penning traps that integrate cylindrical NdFeB permanent magnets and trap electrodes in unitary architectures.  These traps are used to capture and store highly charged ions extracted from the NIST EBIT source, in particular charge states of neon, argon, krypton, and nitrogen.  Ion storage times of order 1 second, limited primarily by background gas pressure, are demonstrated in both a one-magnet trap designed for simplicity of construction and in a two-magnet trap designed for field homogeneity and optical access to the trapping region.  The unitary architecture could find various applications in compact instrument development, for example as portable mass analyzers.

Our ongoing work at NIST focuses on ion studies in the two-magnet Penning trap.  Measurements of collisions, charge exchange rates, effective ion temperatures, and cooling rates are possible in the existing apparatus.  Base pressure as low as 1.0 $\times 10^{-7}$ Pa ($7.8 \times 10^{-10}$ Torr) has been obtained at room temperature, yielding ion storage times up to 3.8 seconds.  A precision leak valve and gas injector nozzle allow for controlled modification of the background gas pressure in the trapping chamber.  An upcoming publication will provide details on the capture process and energy distribution of ions extracted from the EBIT.  

The demonstrated ion storage times of $>$1 second are sufficient for various spectroscopic measurements of interest to astrophysics and plasma diagnostics.  One application particularly suited to this apparatus is measurement of metastable lifetimes.  Well-established efforts at NIST and other EBIT facilities have used interference filters or monochromators to detect fluorescence emitted by ions decaying from metastable levels \cite{Trabert2008}, e.g.  observing the magnetic dipole decay from ArXIV 2p $^2 P_{3/2}$ to 2p $^2 P_{1/2}$ , where the metastable 2p $^2 P_{3/2}$ level is populated during ion production in the hot EBIT plasma \cite{Serpa1998, Lapierre2006}.  Most such measurements in the past have been performed inside an EBIT, producing highly charged ions and then turning off the electron beam long enough to measure fluorescence.  However, since the extraction time of ions from the NIST EBIT is of order 10 $\mu$s, lifetimes of milliseconds or longer may also be observed by detecting fluorescence from ions recaptured in our compact Penning trap.  The two-magnet trap beamline (Fig.  \ref{fig:ExptBeamline}b) includes a lens system with filter and photomultiplier tube for this purpose.  Working with extracted ions in the controlled Penning trap environment allows direct study of certain systematics; for example, pressure dependence of the decay lifetime may be explicitly measured.  Potential disadvantages include a limited detection solid angle and the possibility of new systematic effects due to trapped ion cloud dynamics.  To increase detection solid angle and enable a second systematic check, we are investigating an alternative radiofrequency (RF) ion trap with electrodes that contain large slits for optical access.  The RF trap (currently functioning as an Einzel lens) is located in the experiment beamline (Fig.  \ref{fig:ExptBeamline}b) between the two-magnet Penning trap and TOF detector; it could be loaded either directly from the EBIT or following an initial ion capture stage in the Penning trap.

Longer term applications are motivated by efforts at NIST to ``engineer'' ion states that are not readily produced in the EBIT.  Recent proposals involve producing hydrogenlike ions in circular Rydberg states, for precision spectroscopy to test theory and measure fundamental constants \cite{Jentschura2008, Jentschura2010, Tan2011}.  We have now demonstrated an initial step in realizing this proposal---the capture of bare nuclei (Ne$^{10+}$) in a suitable ion trap.  One possibility for attaching electrons with high angular momentum is charge-exchange with Rydberg atoms \cite{Storry2004}; an atomic beam could be introduced through one of the remaining holes in the two-magnet Penning trap.

For related work with low-Z ions, for which the full energy of the NIST EBIT is unnecessary, the NdFeB Penning traps may also be integrated into a new apparatus with a gas injector and small electron gun, enabling similar experiments at greatly reduced operating cost.  Compact HCI sources based on permanent magnets have been demonstrated at several other facilities; designs range from early ``warm'' EBIT designs utilizing two rings of SmCo$_5$ \cite{Ovsyannikov1999} or NdFeB \cite{Motohashi2000}, to a recent low-energy EBIT featuring 40 NdFeB and soft iron elements \cite{Xiao2012}.

\begin{acknowledgement}
This research was performed while one author (NDG) held a National Research Council Research Associateship Award at NIST.
\end{acknowledgement}

\bibliography{NDGCDAMOP}
\end{document}